\documentclass[fleqn,usenatbib]{mnras}

\usepackage{newtxtext,newtxmath}

\usepackage[T1]{fontenc}
\usepackage{ae,aecompl}

\usepackage{lipsum}
\usepackage{xcolor}

\usepackage{graphicx}	
\usepackage{amsmath}	
\usepackage{amssymb}	

\definecolor{light}{rgb}{0.6,0.6,0.7}

\newcommand{\tdisk }{t_{\rm disk}}

\newcommand{\Mearth}{M_\oplus}
\newcommand{\Rearth}{R_\oplus}
\newcommand{\Msun  }{M_\odot}

\newcommand{\Rp    }{R_{\rm p}}

\newcommand{\yr}{{\rm yr}}

\graphicspath{{./}{plots/}}

\newcommand{\edit}[1]{{\color{red}#1}}

\title[Formation of short-period planets]{Formation of short-period planets by disk migration}

\author[Carrera et al.]{
Daniel Carrera,$^{1}$\thanks{E-mail: danielc@psu.edu, dcarrera@gmail.org}
Eric B. Ford,$^{1,2}$
Andre Izidoro$^{3}$
\\
$^{1}$Department of Astronomy and Astrophysics, 525 Davey Laboratory, The Pennsylvania State University, University Park, PA 16802, USA\\
$^{2}$Institute for CyberScience, The Pennsylvania State University, University Park, PA, 16802, USA\\
$^{3}$UNESP, Univ. Estadual Paulista - Grupo de Din\^amica Orbital \& Planetologia, Guaratinguet\'a, CEP 12516-410 S\~ao Paulo, Brazil
}

\date{Accepted XXX. Received YYY; in original form ZZZ}

\pubyear{2019}

\begin{document}
\label{firstpage}
\pagerange{\pageref{firstpage}--\pageref{lastpage}}
\maketitle

%
%
\begin{abstract}
Protoplanetary disks are thought to be truncated at orbital periods of around 10 days. Therefore, origin of rocky short period planets with $P < 10$ days is a puzzle. We propose that many of these planets may form through the Type-I migration of planets locked into a chain of mutual mean motion resonances.
We ran N-body simulations of planetary embryos embedded in a protoplanetary disk. The embryos experienced gravitational scatterings, collisions, disk torques, and dampening of orbital eccentricity and inclination. We then modelled Kepler observations of these planets using a forward model of both the transit probability and the detection efficiency of the Kepler pipeline. 
We found that planets become locked into long chains of mean motion resonances that migrate in unison. When the chain reaches the edge of the disk, the inner planets are pushed past the edge due to the disk torques acting on the planets farther out in the chain. Our simulated systems successfully reproduce the observed period distribution of short period Kepler planets between 1 and 2 $\Rearth$. However, we obtain fewer closely packed short period planets than in the Kepler sample.
Our results provide valuable insight into the planet formation process, and suggests that resonance locks, migration, and dynamical instabilities play important roles the the formation and evolution of close-in small exoplanets.
\end{abstract}
%
%

%
%
\section{Introduction}
\label{sec:intro}

An important interaction between a protoplanetary disk and a magnetised star is the disruption of the disk at the Alfv\'en radius \citep[e.g.][]{Long_2005}, where the magnetic energy density due to the star's magnetic field equals the kinetic energy density due to the in-falling material. At the same time, the magnetic connection between the star and the disk causes the star's rotational angular momentum to become ``locked'' at a point near the Alfv\'en radius \citep[e.g.][]{Koenigl_1991,Ostriker_1995,Long_2005}. In other words, the inner edge of a protoplanetary disk is expected to be approximately at the point of co-rotation with the host star,

\begin{equation}
    r_{\rm in}  \approx r_{\rm co}
                = \sqrt[\uproot{15} 3]{\frac{G M_\star}{\Omega_\star^2}}.
\end{equation}

\begin{figure}
  \centering
  \includegraphics[width=0.47\textwidth]{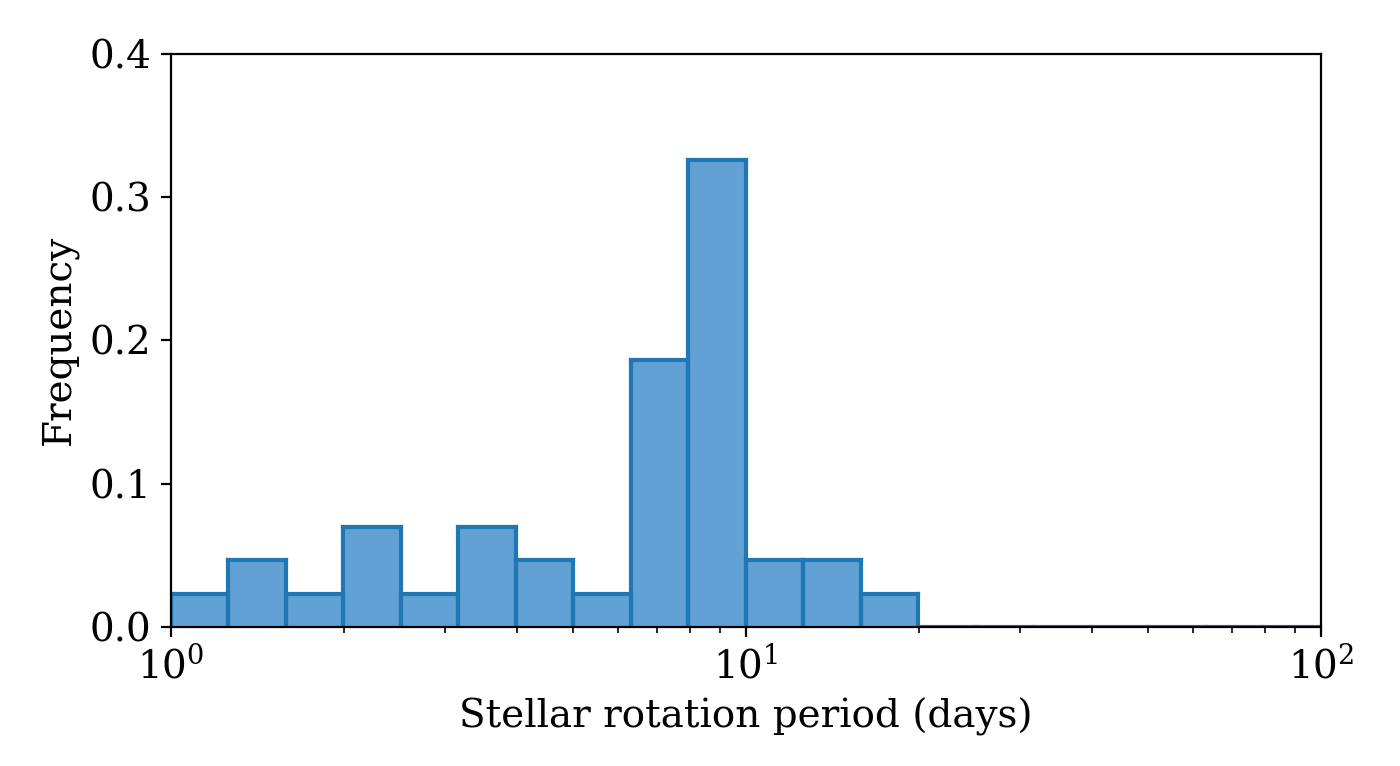}\\
  \includegraphics[width=0.47\textwidth]{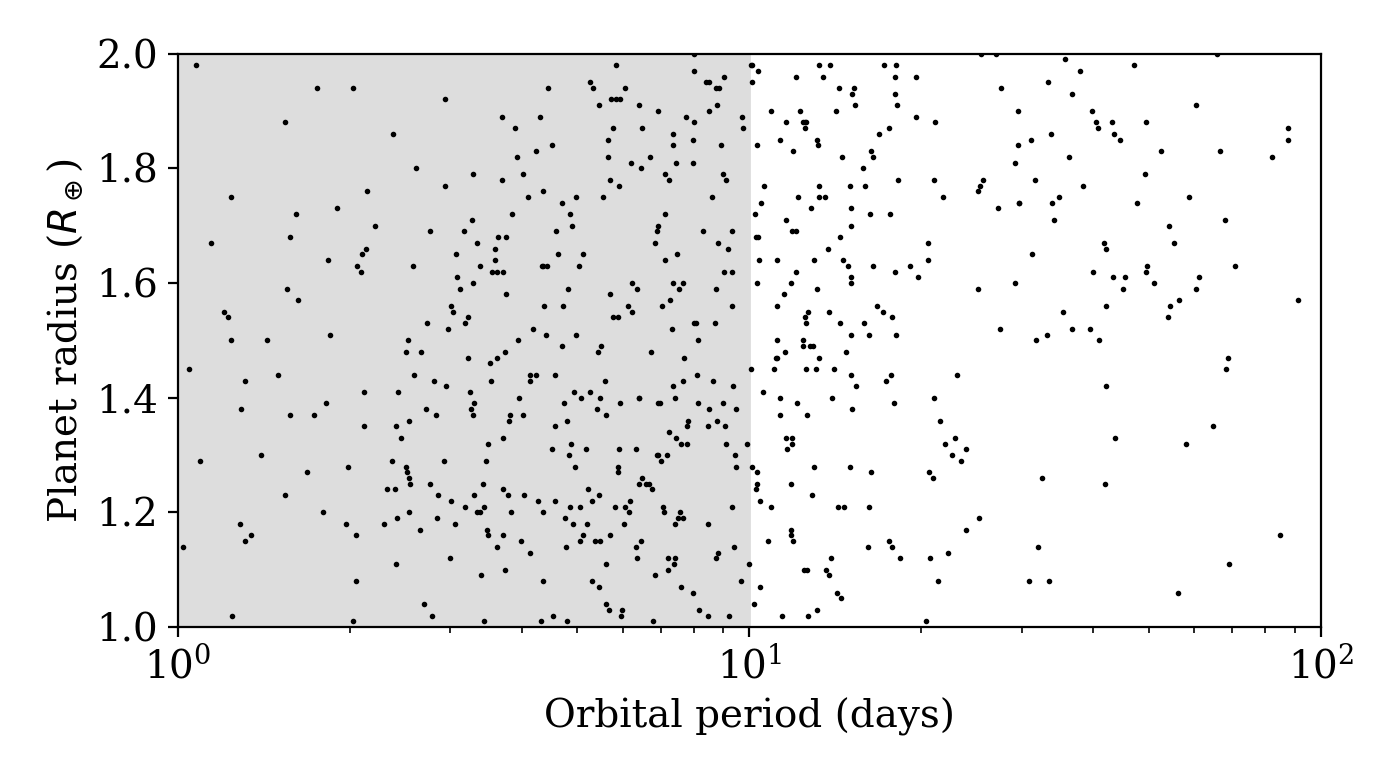}\\
  \includegraphics[width=0.47\textwidth]{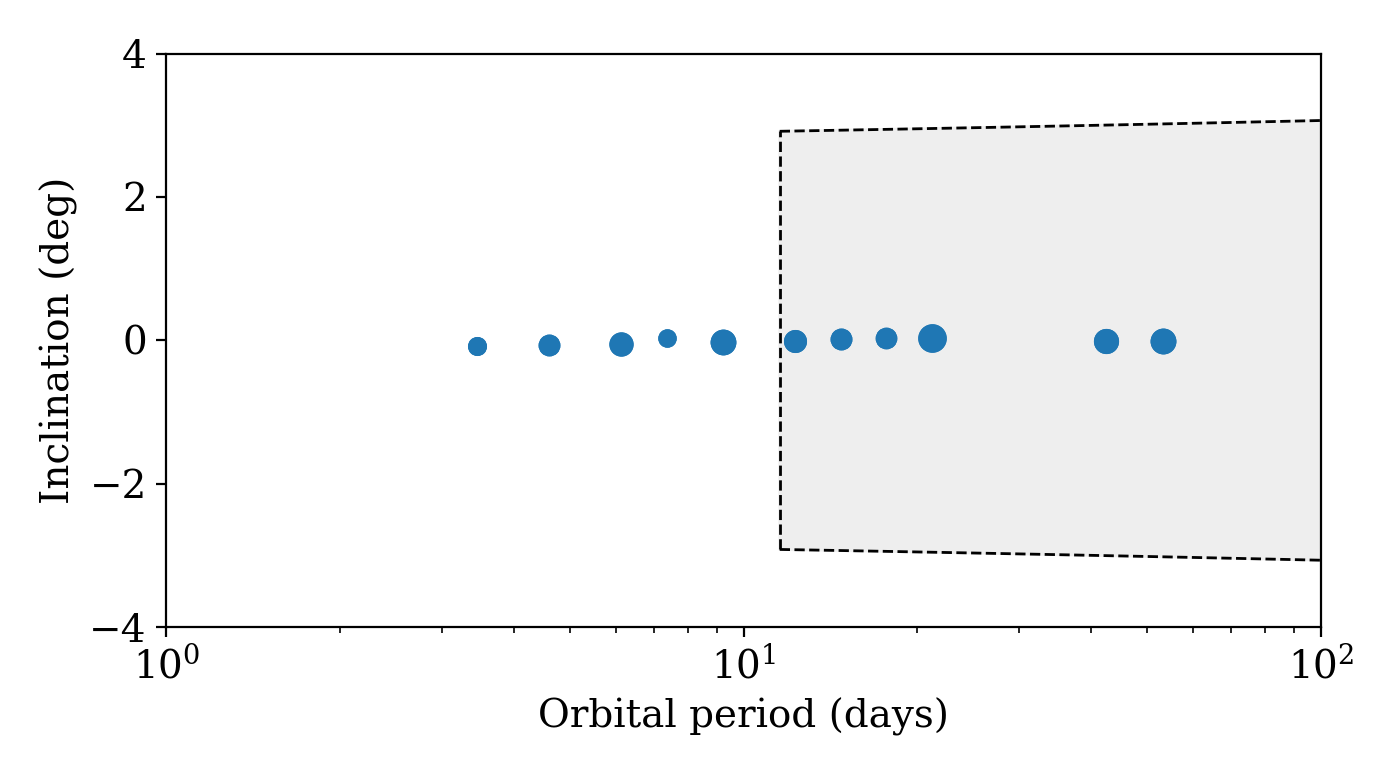}
  \caption{Rotational periods of 433 young FGK stars in the Orion Nebula (top), showing a sharp peak at $\sim$10 days \citep{Herbst_2002}. The disk is thought to stop at the co-rotation point. Yet, the Kepler catalogue (middle) contains many super-Earths interior to 10 days. We propose that these short period planets may form when a group of protoplanets become locked into a chain of mutual mean motion resonances (bottom) which then migrates past the inner edge of the disk.}
  \label{fig:intro}
\end{figure}

Figure \ref{fig:intro} shows the distribution of rotation periods of 433 young FGK stars in the Orion Nebula \citep{Herbst_2002}, showing that most of these stars have rotation periods of around 10 days. Therefore, it is likely that protoplanetary disks are usually truncated at around 10 days. This estimate is also consistent with Keck interferometric observations of FU Orionis, which suggest a disk inner edge at $0.07 \pm 0.02$ AU, or at around 10 days \citep{Eisner_2011}. Indeed, the occurrence rate of sub-Neptunes appears to drop sharply for $P < 10$ days \citep[e.g.][]{Youdin_2011,Mulders_2015,Petigura_2018,Hsu_2019}. How these planets formed, and how they obtained those short orbital periods, is an interesting puzzle, and is the subject of this investigation.

For a single sub-Neptune planet experiencing Type-I migration, disk torques are expected to simply deposit the planet at the inner edge of the disk. This can occur because of the loss of Lindblad torques just inside the disk \citep{Goldreich_1980}, or a large co-rotation torque or the reflection of waves may halt migration at or just outside the inner edge of the disk \citep[e.g.][]{Tanaka_2002,Masset_2006,Terquem_2007,Tsang_2011}. In a recent work, \citet{Lee_2017} argued that this feature disqualifies Type-I migration as a mechanism for forming short period planets, as the period distributions that they obtained did not match observations.

In this work we show that this simple view of migration is not a complete story. Sub-Neptune planets typically form in multiple planet systems where dynamical effects between planets are important \citep[e.g.][]{Terquem_2007}. In this work we show that the combination of Type-I migration with resonance locks of multiple sub-Neptune planets, not only produces short period planets, but also predicts a period distribution that is consistent with that observed in the Kepler field.

Planet formation is thought to begin with the formation of small $1 - 100$ km bodies called planetesimals \citep{Chiang_2010,Johansen_2014}. Planetesimals first experience super-exponential runaway growth \citep{Greenberg_1978,Wetherill_1989,Kokubo_1996} followed by a period of slower oligarchic growth, which ends when the oligarchs reach their isolation mass \citep{Kokubo_1998,Kokubo_2000,Thommes_2003,Chambers_2006}. The formation of oligarchs may overlap with, or be followed by, pebble accretion. Pebble accretion is likely to be particularly important for the formation of giant planet cores at large orbital separations \citep{Lambrechts_2012,Johansen_2017}. Once the planet reaches a certain size, the Lindblad and co-rotation torques cause the planet to migrate inward \citep{Paardekooper_2010}. In the process of migration, the planets capture each other into chains of mean motion resonances with several planets each \citep{Terquem_2007}. While the disk is present, the resonant chains tend to be stabilised by eccentricity and inclination damping from the disk. After the disk dissipates, these resonant chains often become dynamically unstable which leads to planet mergers and a dynamically hotter system \citep{Cossou_2014,Izidoro_2017}.

Unlike the single planet case, a resonant chain typically does not stop at the edge of the disk, as the outer planets continue to experience inward torques which force the entire chain inward. Supersonic corrections to the classical type I migration torques predict that a planet at the inner edge of the disk may be pushed into the cavity through resonant interaction with planets farther out \citep{Brasser_2018}. Figure \ref{fig:intro} illustrates an extension of that idea to an extended resonant chain. A similar result was also seen in Figure 3 of \citep{Carrera_2018}. In this paper we investigate whether the ``\textit{migrating resonant chain}'' mechanism could be responsible for the majority of the observed short period planet population.

\subsection{Transit selection effects}
\label{sec:intro:obs}

A key part of this investigation is an accurate model of detection biases in transit surveys in general, and the Kepler catalogue in particular. Planets with shorter periods have a higher transit probability, and the detection efficiency of transit surveys depends on the transit depth, number of transits available, and stellar noise. Most authors that tackle this problem attempt to ``reverse'' the process to obtain planet occurrence rates \citep{Howard_2012,Petigura_2013a,Petigura_2013b,Fressin_2013}.

Here we take a different approach. Rather than relying on reverse-engineered occurrence rates, we forward model the selection effects in the Kepler catalogue. In other words, we take the planetary systems that come out our simulations and we apply selection effects and compare that directly against the Kepler catalogue. This approach has many advantages. First, there is a large variance in the occurrence rates reported in the literature and often suggest very different size and period dependence. A review of some of these differences was presented in the SAG 13 report\footnote{\url{https://exoplanets.nasa.gov/exep/exopag/sag/\#sag13}}. These differences arise due to different assumptions regarding completeness, whether they account for false positives, and even variations in the statistical methodology. Choosing among the published occurrence rates can lead to different results. But while inferred occurrence rates vary, the final Kepler catalogue does not. So we go directly to the Kepler catalogue for our data comparison, eliminating a middle step (occurrence rates) that is uncertain and out of our control. An additional benefit of a forward modelling is the ability to use the latest Kepler catalogue (DR25), which is also the most robust, the most homogeneous, and best understood catalogue in terms of completeness.

Finally, some of our analysis relies on the joint detection probability of two planets in the same system in order to obtain a distribution of period ratios (Section \ref{sec:results:period_ratios}). This type of information simply is not available in any occurrence rate because it depends on the mutual inclinations between planet orbits --- two planets with a high mutual inclination will almost never be detected together regardless of their individual detection probabilities. This type of geometric effect can and is included in our forward model.

This paper is organised as follows. In Section \ref{sec:methods} we describe our planet formation model, initial conditions, as well as our forward model of Kepler observations. In Section \ref{sec:results} we present our results. We discuss the implications of our results in Section \ref{sec:discussion}, and conclude in Section \ref{sec:conclusion}.

%
%
\section{Methods}
\label{sec:methods}

\subsection{Formation model}
\label{sec:methods:model}

We use some of the planet formation simulations published in \citet{Carrera_2018}. A full description of the planet formation model is included in this publication. We run N-body simulations using the \textsc{mercury} code with the hybrid integrator \citep{Chambers_1999} and a user-defined force to compute disk torques. The formulas for the disk torques are included in Appendix \ref{app:torques}, and were implemented into \textsc{mercury} by \citet{Izidoro_2017}. Here we simply note that, in Type-I migration, there are two sources of disk torques acting on a planet: Lindblad torque, and co-rotation torque,

\begin{equation}
    \Gamma_{\rm tot} = \Gamma_{\rm Lind} + \Gamma_{\rm CoRot}.
\end{equation}

\edit{For typical disk parameters,} Lindblad torque is negative and co-rotation torque is positive. Lindblad torque normally exceeds the co-rotation torque so that migration is usually inward. But outward migration is also possible. For a steady-state accretion disk, where the negative density and temperature gradients add up to 3/2, outward migration requires $d \ln T / d \ln r < - 0.87$ \citep{Brasser_2017}.


In all simulations the stellar mass is 1 $\Msun$ and we use the protoplanetary disk model of \citet{Bitsch_2015}. The model provides formulas for the disk temperature as a function of metallicity ($Z$) and accretion rate $\dot{M}_{\rm disk}$. As in \citet{Carrera_2018}, we tie the accretion rate to the age of the disk $\tdisk$ using

\begin{equation}\label{eqn:Mdot}
	\log_{\rm 10}\left( \frac{\dot{M}_{\rm disk}}{M_\odot / \yr} \right)
    = - 8 - 1.4\,\log_{\rm 10}\left( \frac{\tdisk + 10^5 \yr}{10^6 \yr} \right).
\end{equation}

In this formula, $\tdisk = 0$ corresponds to a stellar age of $t_\star = 10^5$ yr, which roughly corresponds to the moment when the disk becomes gravitationally stable \citep{Bitsch_2015}. Our simulations all begin at $\tdisk = 1$ Myr, which roughly corresponds to the timescale of embryo formation at 1 AU \citep{Kokubo_2000}.

Given the disk temperature and $\dot{M}_{\rm disk}$, we use \citet{Shakura_1973} $\alpha$-viscosity and hydrostatic equilibrium to solve for the gas surface density $\Sigma$

\begin{eqnarray}
	\dot{M}_{\rm disk} &=& 3 \pi \alpha H^2 \Omega \Sigma,\\
	T &=& \left( \frac{H}{r} \right)^2 \frac{G \Msun}{r} \frac{\mu}{\mathcal{R}},
\end{eqnarray}
where $G$ is the gravitational constant, $\mu = 2.3$ is the molecular mass, and $\mathcal{R}$ is the gas constant.

Finally, we assume that disk photoevaporation becomes an important process at $\tdisk = 5$ Myr and it dissipates the disk quickly \citep[e.g.][]{Gorti_2009}. To simulate that, we fix the disk temperature and reduce its mass exponentially with an e-folding timescale of $10^4$ yr for 10 e-folds. At 5.1\,Myr, we remove the disk entirely, for computational convenience.

\subsection{Initial conditions}
\label{sec:methods:init}

We assume that planetesimals form early, and do not experience significant migration either due to aerodynamic drag or to disk torques. In other words, the surface density of embryos reflects the solid surface density of the disk at $\tdisk \approx 0$. Therefore we set

\begin{equation}
	\Sigma_{\rm solid} = Z \; \Sigma_{\rm gas},
\end{equation}
where $\Sigma_{\rm gas}$ is computed at $\tdisk = 0$ (Equation \ref{eqn:Mdot}). The full list of simulations is shown in Table \ref{tab:models}. We vary the disk metallicity across $Z$ = 0.5\%, 1\%, and 2\%. For each metallicity we run 200 simulations. In each simulation the innermost embryo is placed at 1 AU. We divide the disk into 125 radial bins starting at 1 AU, such that each bin has a solid mass of 0.4 $\Mearth$, and we place a 0.4 $\Mearth$ embryo in the middle of each bin. This works out to a dynamical separation of 7.3 mutual Hill radii for the first two embryos in our baseline model ({\tt Z01}). As the semimajor axes increase, the dynamical separations drop down to 0.4 $R_{\rm Hill}$. Classical models of oligarchic growth predict embryo separations of 5-10 $R_{\rm Hill}$ \citep[e.g.][]{Kokubo_2000}. Therefore, these simulations effectively model the final stages of embryo formation.

The total solid mass available to make planets is 50 $\Mearth$. All embryos are given an initial eccentricity of $e = 0.002$ and inclination $I = 0.10^\circ$, as well as random mean anomaly, argument of preicenter, and longitude of ascending mode, chosen uniformly between $0^\circ$ and $360^\circ$.

\begin{table}
  \caption{We ran planet formation simulations for three values of the disk metallicity $Z$. All models had a stellar mass of $1 \Msun$ and 125 embryos with a mass of $0.4 \Mearth$ each. The embryos start at 1 au and the separations are chosen to follow the solid surface density $\Sigma_{\rm solid} = Z\,\Sigma_{\rm gas}$. The disk metallicity affects the initial location of embryos $a_0$, as well as the disk opacity, temperature, and the strength and direction of disk torques.}
  \label{tab:models}
  \begin{tabular}{lcccccc}
  Model & $Z$ & $N_{\rm emb}$ & $M_{\rm emb}$ & $a_{\rm in}$ & $a_{\rm out}$ & $N_{\rm runs}$\\
  \hline
  \tt{Z05}        & 0.005 & 125 & 0.4$\Mearth$ & 1 AU & 7.6 AU & 200 \\
  \tt{Z10}        & 0.010 & 125 & 0.4$\Mearth$ & 1 AU & 6.0 AU & 200 \\
  \tt{Z20}        & 0.020 & 125 & 0.4$\Mearth$ & 1 AU & 4.9 AU & 200 \\
  \end{tabular}
\end{table}

\subsection{Planet radius}
\label{sec:methods:radius}

Before we can simulate Kepler observations, we need to assign each planet a physical radius, as that determines the transit depth. For short period planets, with periods less than 10 days, we assume that any primordial atmosphere has been fully evaporated and the planet is rock-iron core \citep[e.g.][]{Carrera_2018}. To compute the radius of the rock-iron core we use the planet structure model of \citet{Zeng_2016}. Naturally, the planet radius depends partly on the iron mass fraction, which is unknown for exoplanets and varies widely in the solar system. To account for this source of uncertainty, every time we compute a core radius we randomly pick an iron mass fraction between 20\% and 50\%. This is a slightly wider range than the one observed in the solar system.

Finally, for planets beyond 10 days, the atmosphere may represent a very large fraction of the planet's transit radius. Rather than modelling atmosphere processes, which have many uncertainties, we simply consider two limiting cases: we either treat all $P > 10$ day planets as bare rocky cores, or treat them all as Jupiter-radius planets. This gives us a range from the smallest and largest transit radii that are consistent with each simulation.

\subsection{Simulating transit observations}
\label{sec:methods:obs}

We use the ExoplanetsSysSim\footnote{\url{https://github.com/eford/ExoplanetsSysSim.jl}} package to simulate transit observations of planetary systems by the Kepler spacecraft. The package computes the geometric transit probability, as well as the detection efficiency of the Kepler pipeline taking into account the stellar noise, transit depth, and number of transits. To obtain an accurate simulation of the number of transits, we use the observing window function for each Kepler star, including any gaps in the observations, such as when a star falls onto a bad pixel on the Kepler CCD.

After a simulated planetary system has been selected, and planetary radii have been assigned (Section \ref{sec:methods:radius}), we randomly select a random G-type star from the Kepler input catalog. We put all the planets around that star, preserving all of their orbital properties (semimajor axis, eccentricity, inclination, longitude of ascending node, argument of periastron, and mean anomaly). Preserving the orbital angles allows us to accurately model the joint probability that any given pair of planets both transit.

\begin{figure}
  \begin{flushleft}
  \includegraphics[width=0.42\textwidth]{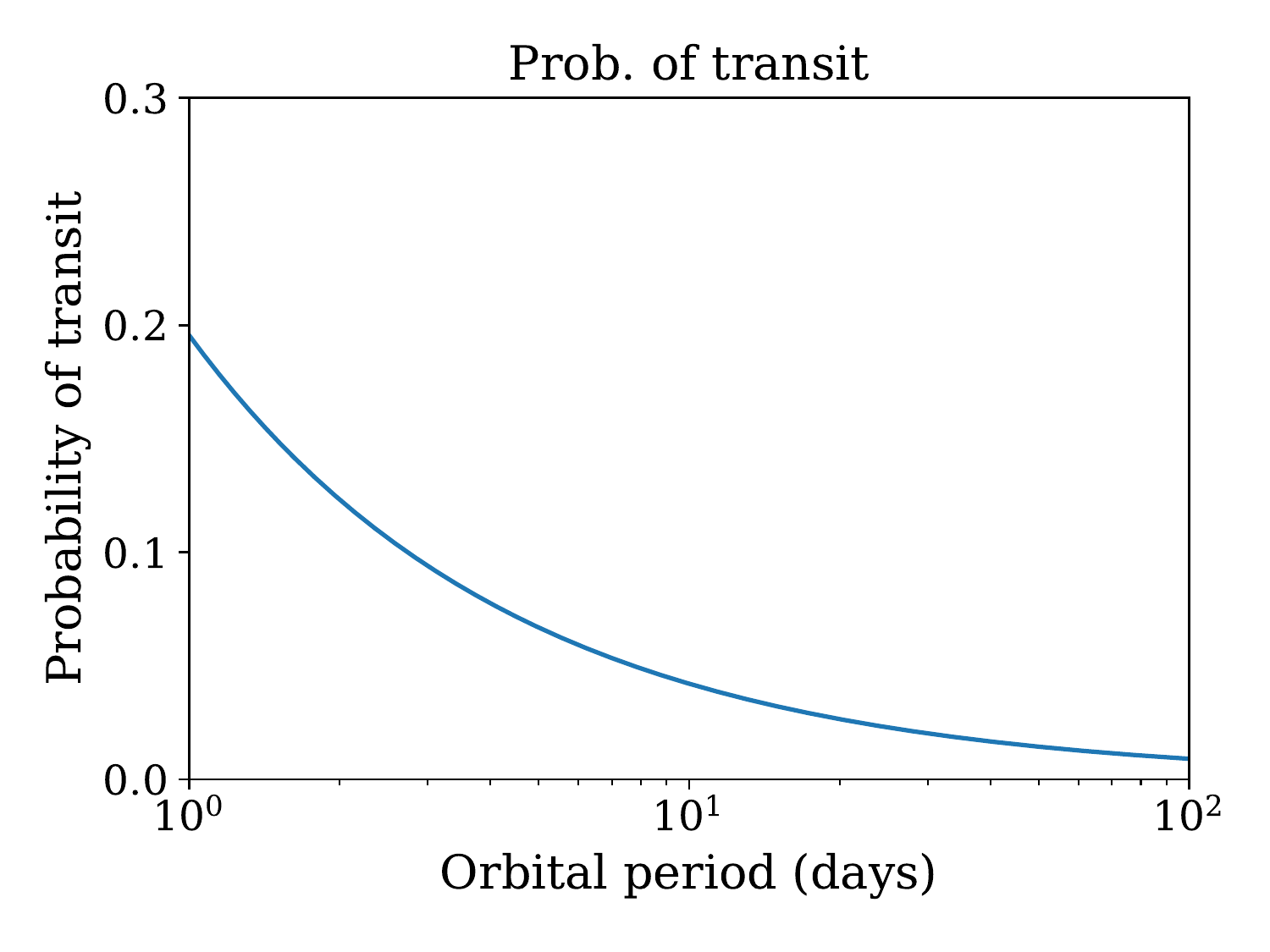}\\
  \includegraphics[width=0.47\textwidth]{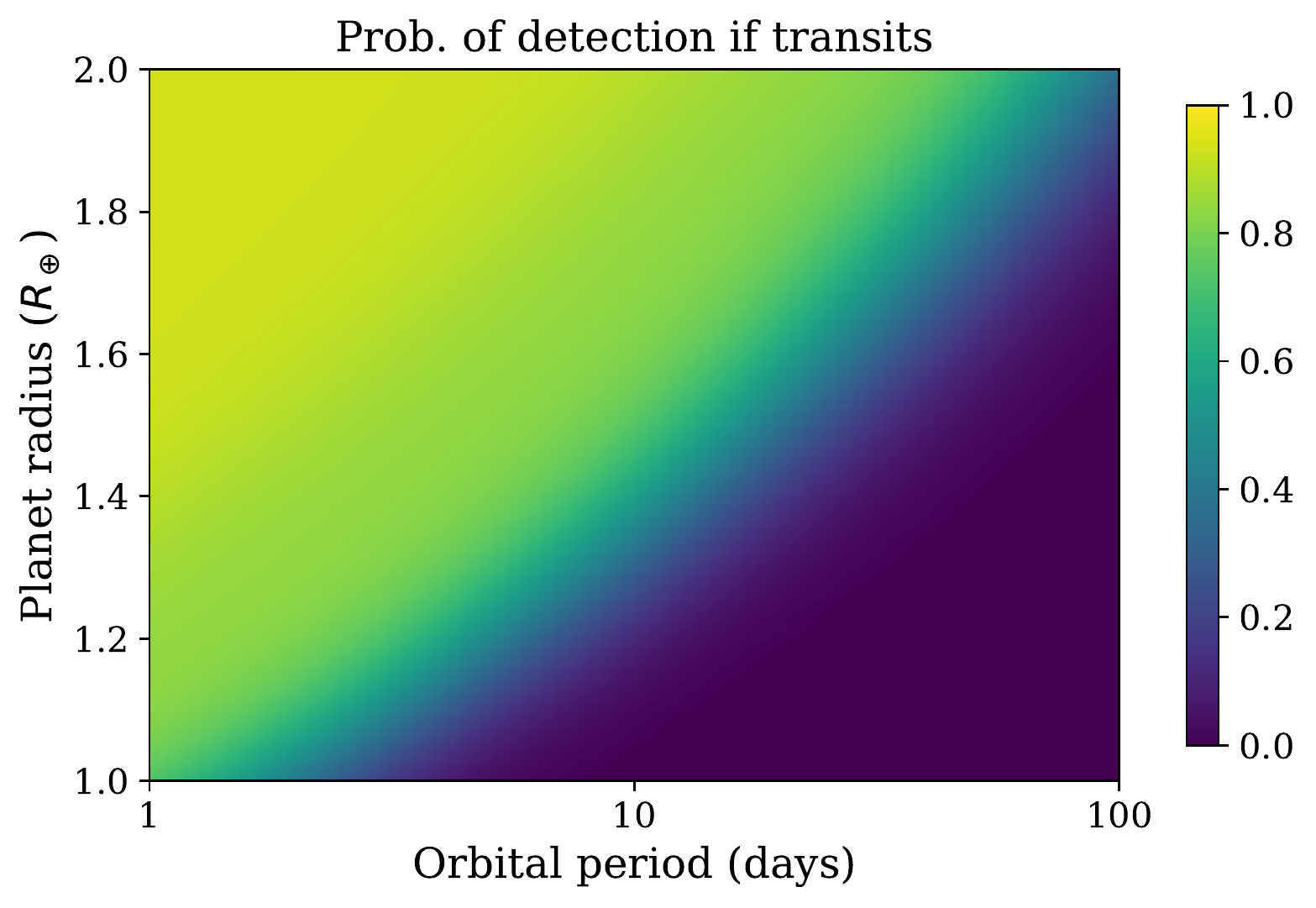}\\
  \includegraphics[width=0.47\textwidth]{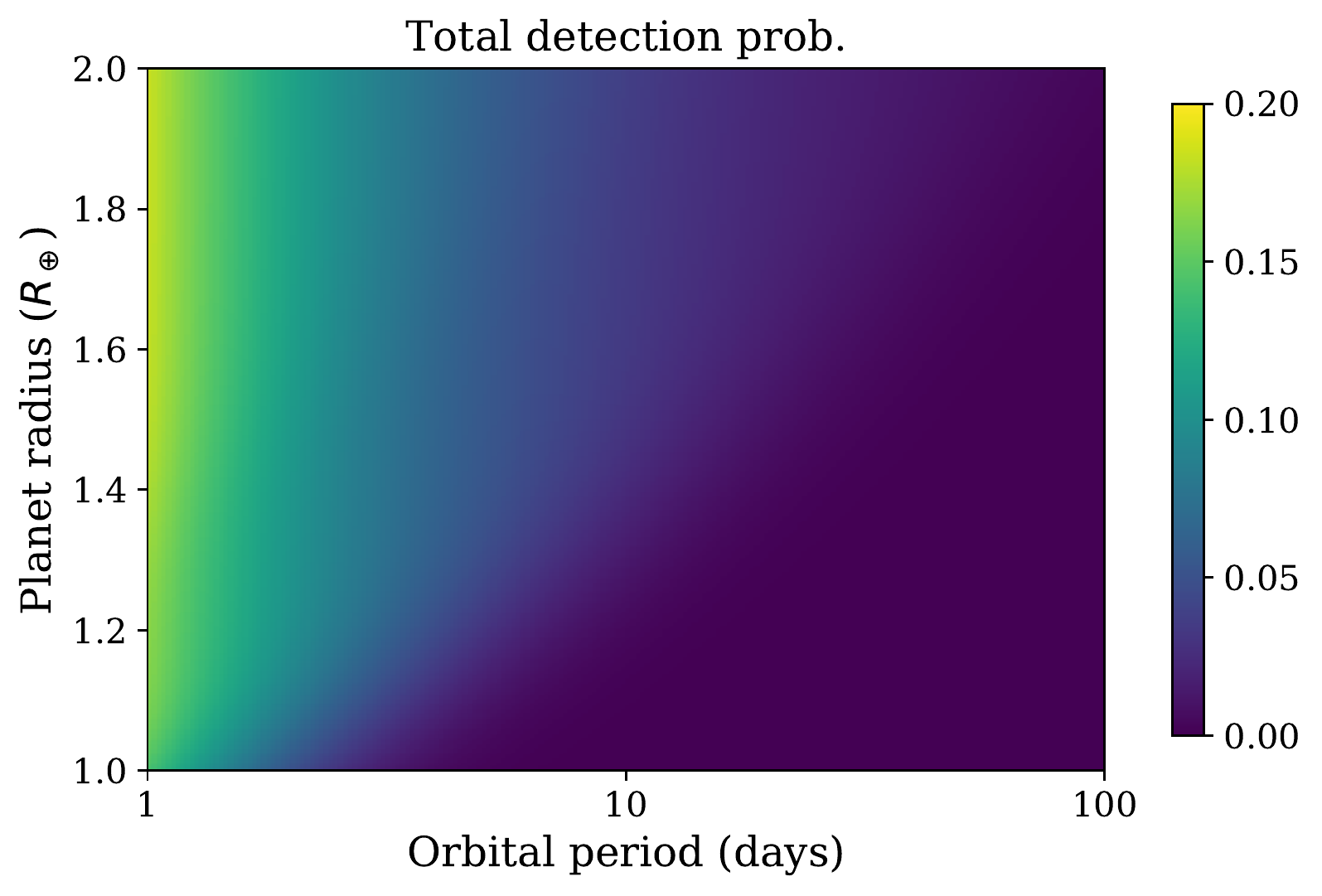}
  \end{flushleft}
  \caption{Kepler detection model for a sample G-dwarf in the Kepler target list. The total probability that a planet is detected (bottom) is the product of the probability that the planet transits (top) and the detection efficiency of the Kepler pipeline (middle). These vary from star to star as they depend on the stellar noise and the observation window for each star. At longer orbital periods, the position of the planet within its orbit significantly affects the number of transits that occur within the observing window for that star, which affects the detection efficiency.}
  \label{fig:syssim}
\end{figure}

To simulate observed occurrence rates (Section \ref{sec:results:periods}), we only need to consider each plant in isolation. Figure \ref{fig:syssim} shows how we compute total probability that a given planet is detected. First, we compute the sky-averaged probability that a planet transits. This quantity is primarily a function of semimajor axis, eccentricity, and stellar radius. Then we compute the probability that, if the planet transits, it would be detected by the Kepler pipeline. This quantity is a function of the transit depth, number of transits, and the amount of stellar noise. The product of these two values gives us a total detection probability, which is proportional to the occurrence rate.

\begin{equation}
    P(A) = P_{\rm tran}(A) \cdot P_{\rm pipe}(A)
\end{equation}

For computing the joint probability that both planets in a planet pair are detected (Section \ref{sec:results:period_ratios}), we need to take into account the relative orientations and shapes of the two planet orbits. Using that information, we compute a sky-averaged probability that both planets in a pair transit. To compute the detection efficiency of the Kepler pipeline, we make the simplifying assumption that the pipeline detection efficiencies are independent, so we compute those values as in the single-planet case. Finally, the total probability of detection is the product of all three values,

\begin{equation}
    P(A,B) = P_{\rm tran}(A,B) \cdot P_{\rm pipe}(A) \cdot P_{\rm pipe}(B)
\end{equation}

\subsection{Bootstrap procedure}
\label{sec:methods:bootstrap}

\begin{figure}
  \centering
  \includegraphics[width=0.47\textwidth]{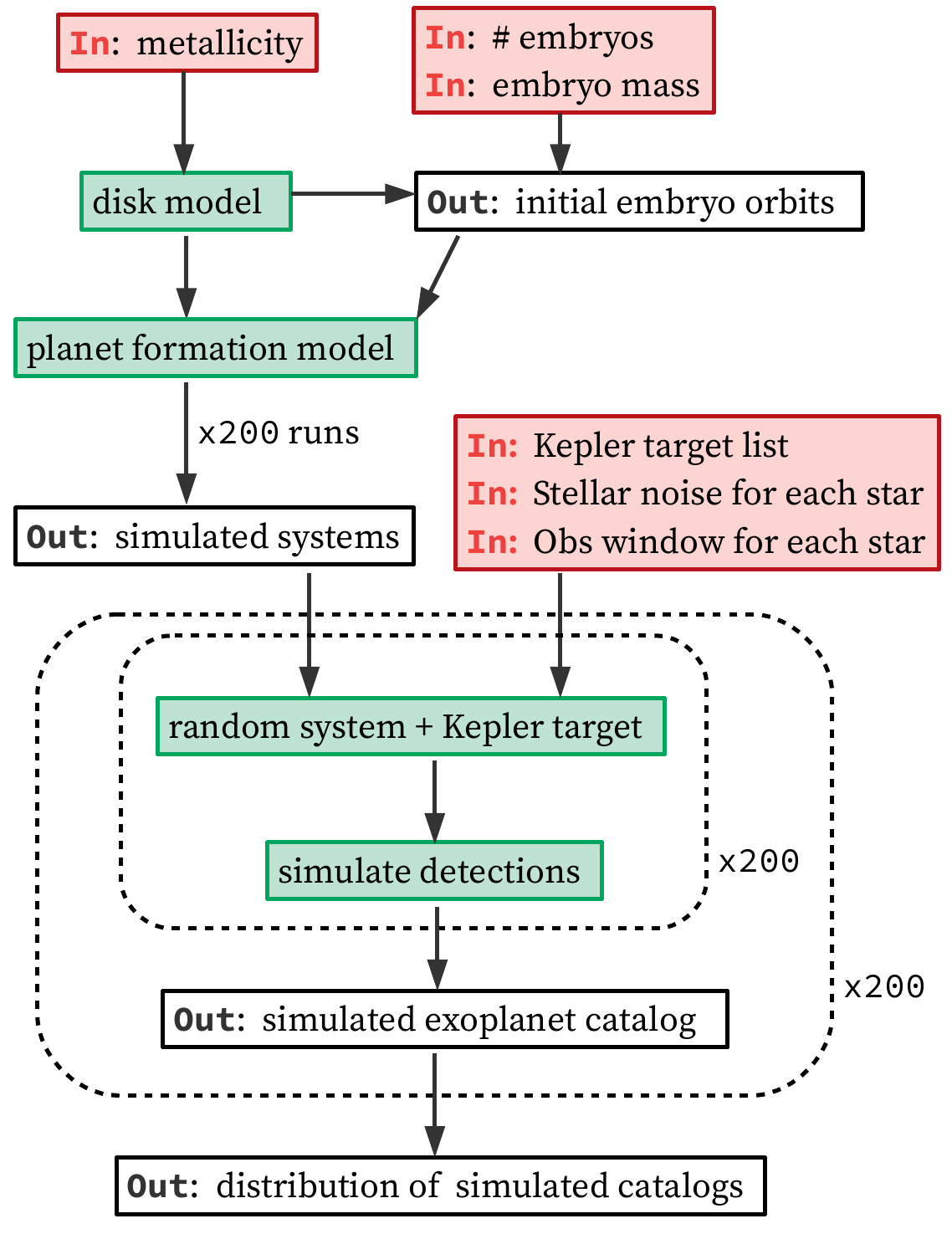}
  \caption{Sketch of our bootstrap method. We run 200 planet formation simulations with the same disk parameters. We randomly select 200 systems (with replacement) and assign each system a random G-type star from the Kepler input catalogue, including the radius, stellar noise, and observation window for that star. We simulate the observations of these systems to produce a simulated exoplanet catalogue. This is 1 bootstrap sample. We draw 200 bootstrap samples to obtain a distribution of simulated exoplanet catalogues that we then compare with the Kepler exoplanet catalogue.}
  \label{fig:diagram}
\end{figure}

If we were to simply apply transit observations (Section \ref{sec:methods:obs}) to the simulated systems, we would obtain estimated occurrence rates, but we would not have confidence intervals on those quantities. To obtain error bars, we use the bootstrap process sketched in Figure \ref{fig:diagram},

\begin{enumerate}
    \item For a given disk model, run 200 simulations to generate 200 simulated planetary systems.
    
    \item Randomly select, with replacement, 200 systems from the list of simulations.
    
    \item Assign to each system a G-type Kepler target list, assign each planet an iron fraction between 20\% and 50\%, and simulate Kepler observations (Sections \ref{sec:methods:radius} and \ref{sec:methods:obs}).
    
    \item Use the simulated exoplanet catalog to compute planet occurrence rates in different period bins (Section \ref{sec:results:periods})
    
    \item Repeat steps 2-4 two hundred times.
\end{enumerate}

Each iteration of steps 2-4 produces a possible exoplanet catalogue and list of occurrence rates. Comparing across all 200 bootstraps samples, we can obtain medians and confidence intervals on all the occurrence rates. The confidence intervals thus produced naturally take into account any uncertainties due to small number statistics that may occur when only a few simulations produce planes in a certain period range. This bootstrap procedures also ensures that each system is selected many times, and is observed many times around different Kepler targets, with slightly different transit depths due to different iron fractions.

While the procedure described above is expressed in terms of single planet occurrence rates, the procedure for computing the frequency of planet pairs and period ratios is entirely analogous.

%
%
\section{Results}
\label{sec:results}

\subsection{Example simulation}
\label{sec:results:example}

\begin{figure*}
  \includegraphics[width=0.95\textwidth]{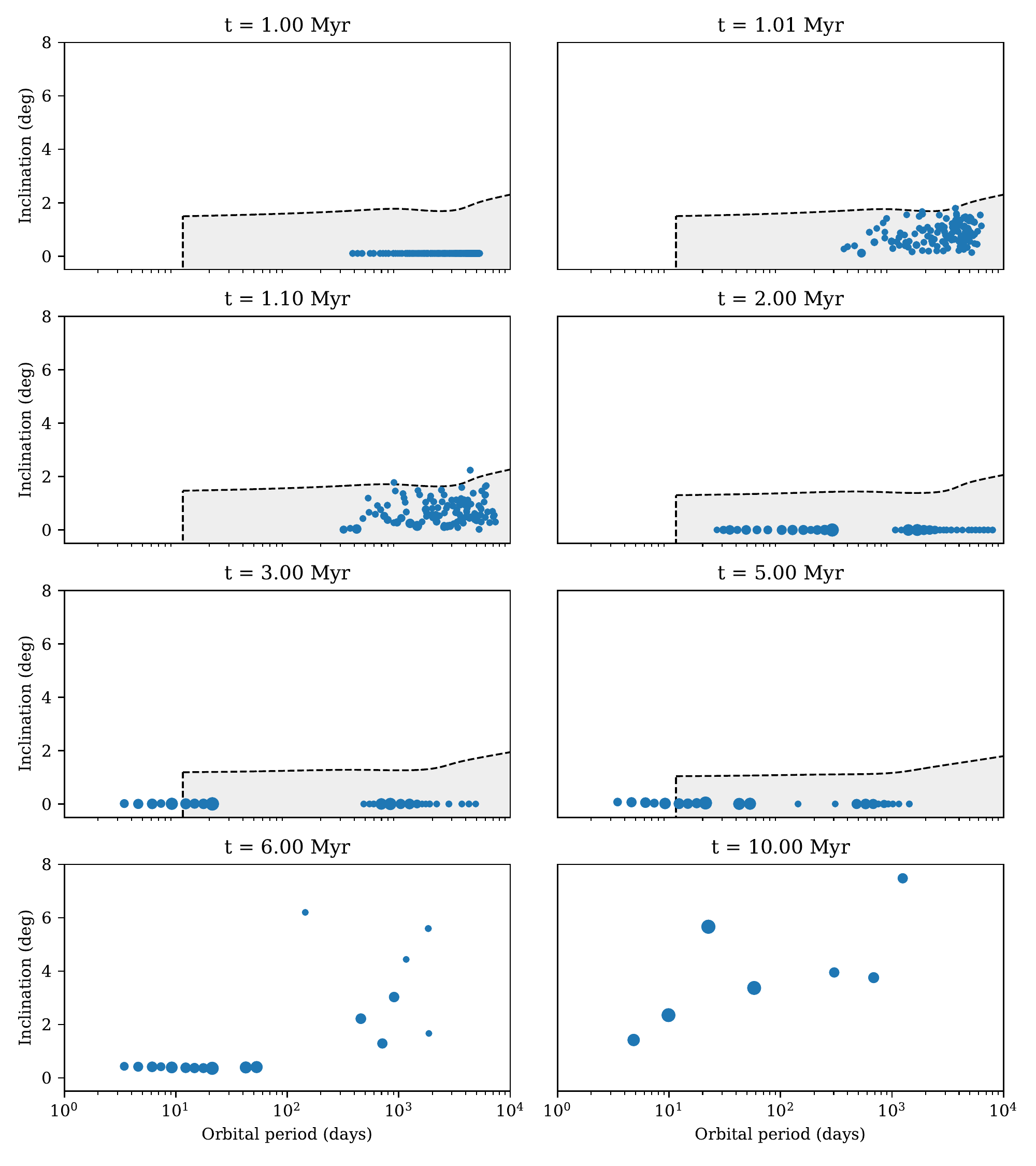}
  \caption{Snapshots from a simulation from Model {\tt Z10}. The protoplanetary disk begins to dissipate at 5 Myr. The grey region shows the scale height of the disk as ${\rm atan}(H/r)$. The run begins with 125 embryos, each with a mass of $0.4 M_\oplus$, at a disk age of 1 Myr. The embryos have strong dynamical interactions that excite their eccentricities and inclinations. As the embryos merge, disk torques dampen eccentricities and inclinations and cause planets to migrate inward. At $t = 2$ Myr the planets have settled into a chain of mean motion resonances, and a migration trap separates the planets into two groups. After 3 Myr the inner chain has pushed past the inner edge of the disk. The snapshot at 5 Myr is right before the disk dissipates. The last two panels show the effect of the dynamical instabilities that occur after the dampening effect of the gas disk has been removed at 5.1 Myr. This run produces two planets with $P < 10$ day.}
  \label{fig:snapshots}
\end{figure*}

Figure \ref{fig:snapshots} shows six snapshots from one of our simulations. The simulation is for a disk with a metallicity of $Z = 1$\% (Model {\tt Z10} in Table \ref{tab:models}). The embryos (blue dots) are initially arranged in near-flat near-circular orbits from 1 to 6 AU. Early on, the embryos experience strong dynamical interactions and collisions, occasionally reaching inclinations as high as the scale height of the disk (grey region). As the embryos grow into planets, dynamical separations increase and torques from the disk dampen the planets' orbital inclinations and eccentricities. This dampening, along with migration, causes the newly forming planets to settle into long chains of mutual mean motion resonances.

The planets locked into a resonant chain migrate together as a group as the disk evolves. Eventually, the innermost planet reaches a zero torque migration region \citep[a planet trap][]{Masset_2006b} at the inner edge of the disk. If the planet were alone, that is where migration would stop, as was done in the model of \citet{Lee_2017}. However, the presence of nearby planets significantly changes the dynamics at this critical junction. The other planets in the chain continue to migrate inward, and in doing so, they force the inner planet past the disk edge and inside the disk inner cavity. This process continues, with more and more planets being pushed past the edge, until it no longer can. At some point the torque on the planets that remain in the disk is no longer sufficient to move the planet chain significantly.

While planets are embedded in the disk, the dampening of eccentricity and inclination helps stabilise the planet orbits. In the bottom two frames of Figure \ref{fig:snapshots} we see what happens to the system after the disk dissipates. With the dampening removed, the resonant chains eventually become dynamically unstable and break apart, leading to a new wave of collisions, and a dynamically hotter system \citep[see][for a detailed discussion]{Izidoro_2017}.

\begin{figure*}
  \centering
  \includegraphics[width=0.95\textwidth]{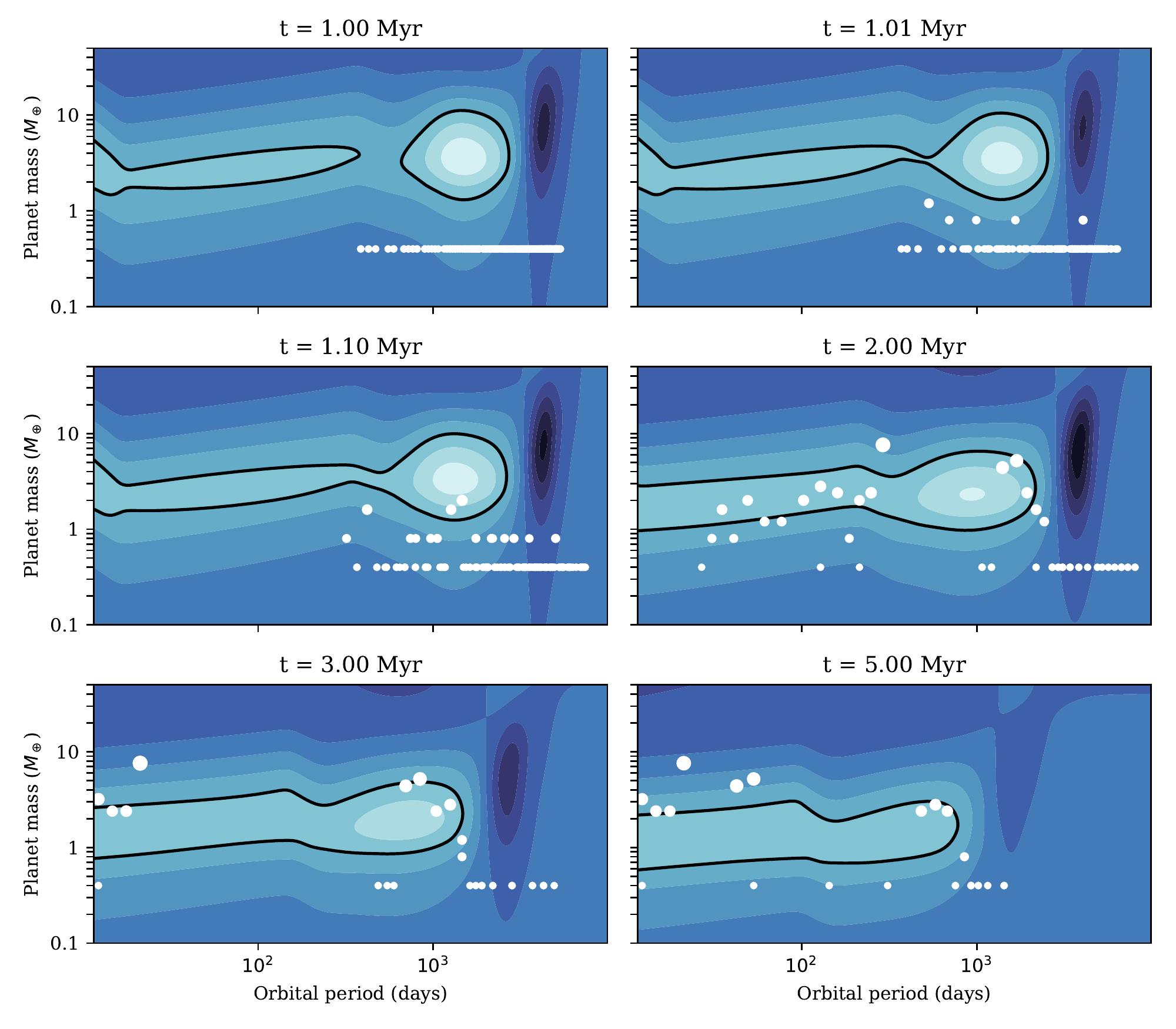}\\
  \includegraphics[width=0.5\textwidth]{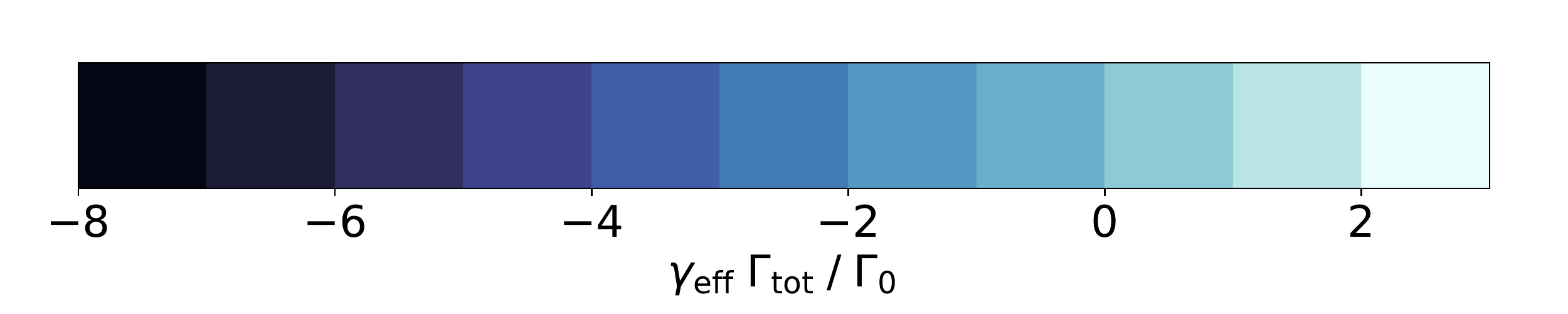}
  \caption{Evolution of the migration map for the protoplanetary disk of Model {\tt Z10}. The disk dissipates after 5.1 Myr. The colorbar shows the torque on the planet, normalised by the effective adiabatic index $\gamma_{\rm eff}$ and $\Gamma_0 = (q/h)^2\,\Sigma\,r^4 \Omega^2$, where $q$ is the planet-star mass ratio, and $h$ is the disk aspect ratio. The black curve marks the zero-torque boundary. A positive torque leads to inward migration, and negative torque leads to outward migration. However, when a group of planets are locked into mean motion resonances, the resonant chain moves as a group, and the direction and speed of migration results from the net torque across all the plants. The white circles represent the planets in the sample simulation shown in Figure \ref{fig:snapshots}.}
  \label{fig:torques}
\end{figure*}

Figure \ref{fig:torques} shows the same simulation from the point of view of the evolving disk structure. The figure shows the net torque experienced by a planet, as a function of the planet's mass and its location in the disk. The colour scale shows the strength and direction of the net torque on the planet, and the region of positive torque (i.e. outward migration) is marked by a black boundary. Changes in the torque map are caused by changes in the thermal structure of the disk. As the disk evolves, it cools, and the steepness of the temperature gradient (which determines the relative strength of the co-rotation torque) also changes \citep{Bitsch_2015}.

As shown in Figure \ref{fig:torques}, inside a few hundred days, at any given point in time the sign of the torque depends mainly on mass, and not so much on period. But as the disk evolves and cools, the mass range that experiences a net outward torque decreases. The figure also shows the somewhat complex nature of migration for a resonant chain. In general, at any one time some planets in the chain experience negative torques and others experience positive torques, and it is the balance of torques on all the planets that determines the direction and speed of migration of the chain as a whole.

\subsection{Planet radii}
\label{sec:results:radii}

\begin{figure}
  \includegraphics[width=0.47\textwidth]{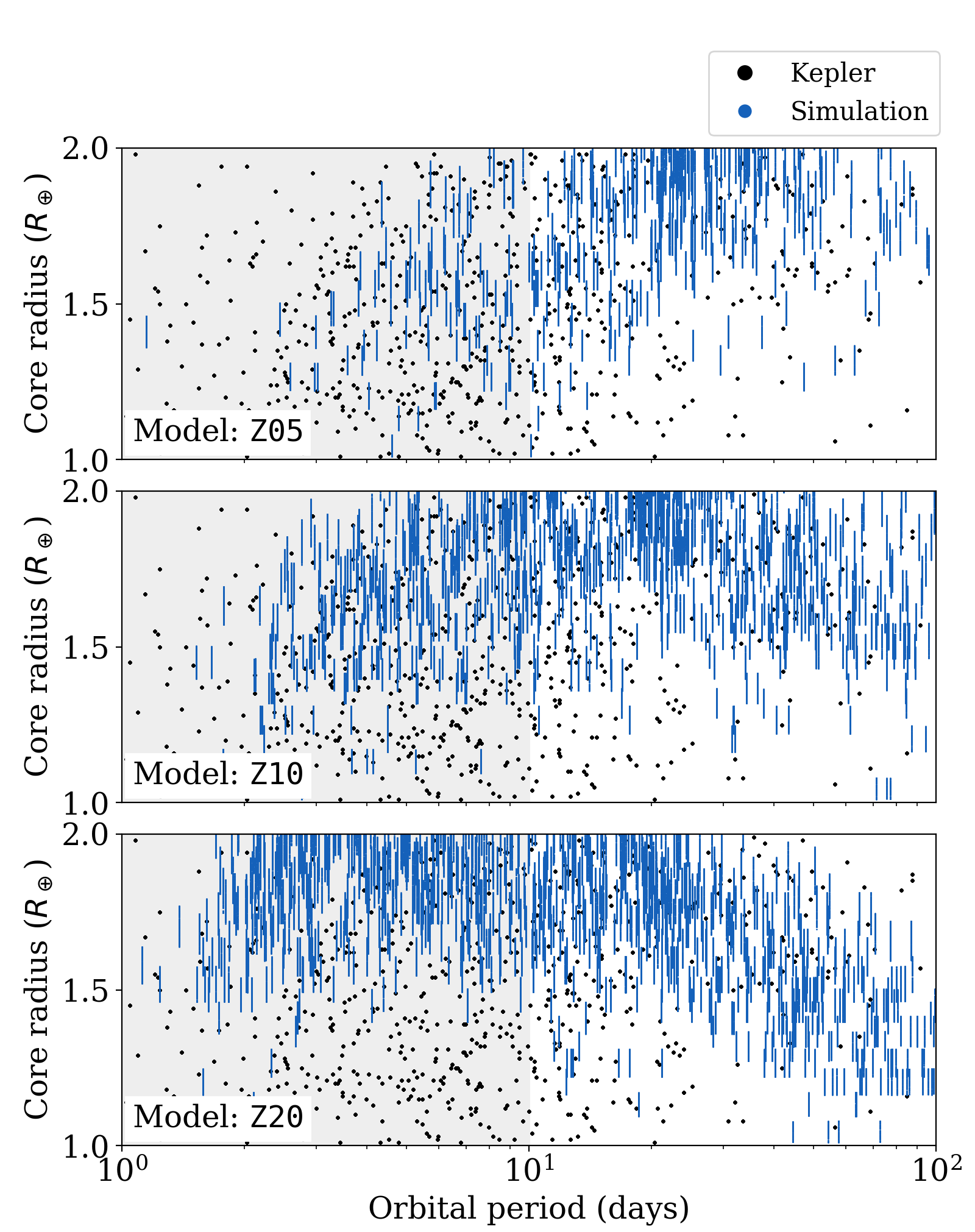}
  \caption{Periods and sizes of Kepler planets (black) and simulated planets (blue) after applying detection biases. For simulated planets, only the size of the rocky core is shown. The vertical lines reflect the fact that each planet is sampled many times, and assigned a random iron fraction of 20-50\%. Models {\tt Z05}, {\tt Z10}, and {\tt Z20} have disk metallicities of $Z = 0.5$\%, 1\%, and 2\% respectively. The grey region marks short period planets, which are likely to be photo-evaporated, while the rest may have gaseous envelopes \citep{Carrera_2018}.}
  \label{fig:R_vs_P}
\end{figure}

Figure \ref{fig:R_vs_P} shows the final periods and core radii of the simulated planets (blue) after taking into account the probability of transit and detection efficiencies of the Kepler pipeline. The vertical lines result from the fact that each planet is sampled many times and each time it receives a random iron fraction, which affects the planet's transit radius. It is worth highlighting that this plot only shows the size of the planet's core, but planets beyond 10 days are likely to have extended gaseous envelopes \citep{Carrera_2018}.

The figure also shows that the disk metallicity clearly has an effect on the final sizes of the planets and the frequency of short-period planets. The simulations with low metallicity disks (Model {\tt Z05}) start out with more sparsely separated embryos that experience fewer collisions and take longer to reach the inner edge of the disk.

\subsection{Period distribution}
\label{sec:results:periods}

\begin{figure}
  \includegraphics[width=0.47\textwidth]{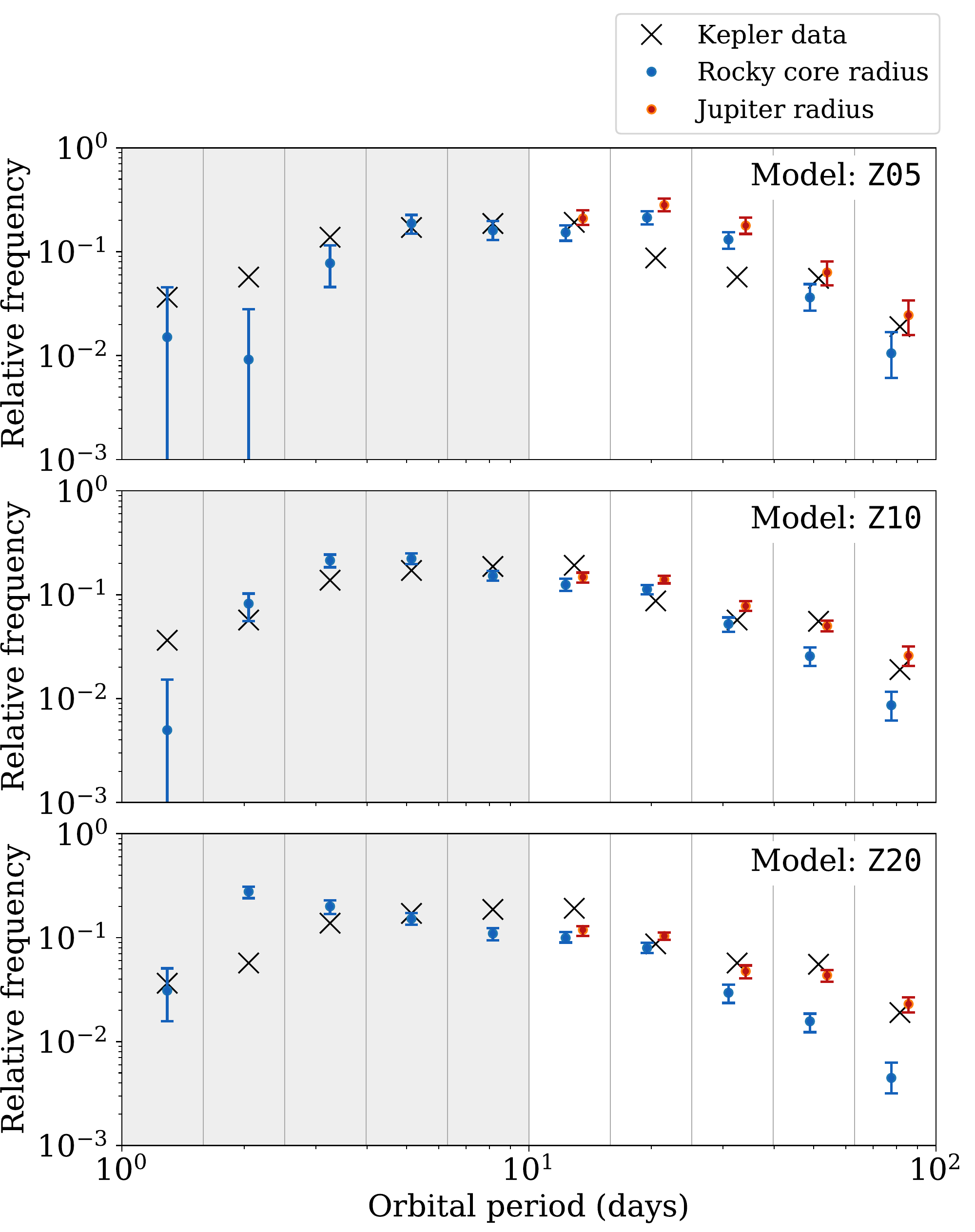}
  \caption{The resonant-chain model (colour) predicts a planet period distribution that is consistent with that observed in the Kepler catalogue (black). The solid point is the median of 200 boostrap resamplings and the error bars cover the middle 90\% interval. The predicted period distribution takes into account the transit probability and the detection efficiency of the Kepler pipeline (Section \ref{sec:methods:obs}). Because planets beyond 10 days may have large envelopes, which increase the planet's radius and detection efficiency, we consider two limiting cases: We either treat all planets as rocky cores (blue) or treat all planets beyond 10 days as Jupiters (red). Both the Kepler and simulated planets received the same size and period cut-offs (see main text).
}
  \label{fig:freq_vs_P}
\end{figure}

One of the most important tests for a formation model for short period planets is its ability to reproduce the observed orbital period distribution. In this investigation we model planet radii and observational effects with the goal of making this particular test as robust as possible for each of the disk models in Table \ref{tab:models}. Figure \ref{fig:freq_vs_P} shows the ``observed'' period distribution of our simulated systems after taking into account observational effects.

For short-period planets ($P < 10$ days) we applied a size cut-off of $\Rp = 1 - 2 \Rearth$ for both Kepler and simulated planets, since our model is geared for the production of small planets. The figure shows that the proposed resonant-chain predicts a short-period planet distribution that is very consistent the one observed in the Kepler field.

The figure also extends the period distribution to orbital periods longer than 10 days. These are certainly not short period planets, and it is likely that many of them possess extended gaseous envelopes \citep{Carrera_2018}. For these planets, modelling the planet atmosphere is fraught with peril. Uncertainty in the planet radius translates into uncertainty in the implied occurrence rates, as larger transit depths are easier to detect by the Kepler pipeline. Instead, our approach is to consider two limiting cases:

\begin{description}
\item[Case A] Treat all planets with $P > 10$ days as rocky cores.

\item[Case B] Treat all planets with $P > 10$ days as Jupiter-radius planets.
\end{description}

This provides the minimum and maximum detection efficiencies for planets beyond 10 days. For planets with $P > 10$ days, the size cut-off is $\Rp > 1 \Rearth$ for both Kepler and simulated planets. In Figure \ref{fig:freq_vs_P}, results assuming that planets are rocky cores are shown in blue, and results assuming Jupiter-radius planets are shown in red. For the low-metallicity model ({\tt Z05}) the rocky-core and Jupiter-radius models are both consistent with the data. For the higher metallicity models ({\tt Z10} and {\tt Z15}), larger radii (red) are favoured in the last two period bins.

Finally, the planets in the Kepler catalogue will certainly have come for a diverse group of protoplanetary disks. A distribution of disk metallicities might reproduce the exoplanet period distribution better than any single disk model alone.

\subsection{Period ratios}
\label{sec:results:period_ratios}

\begin{figure}
  \includegraphics[width=0.47\textwidth]{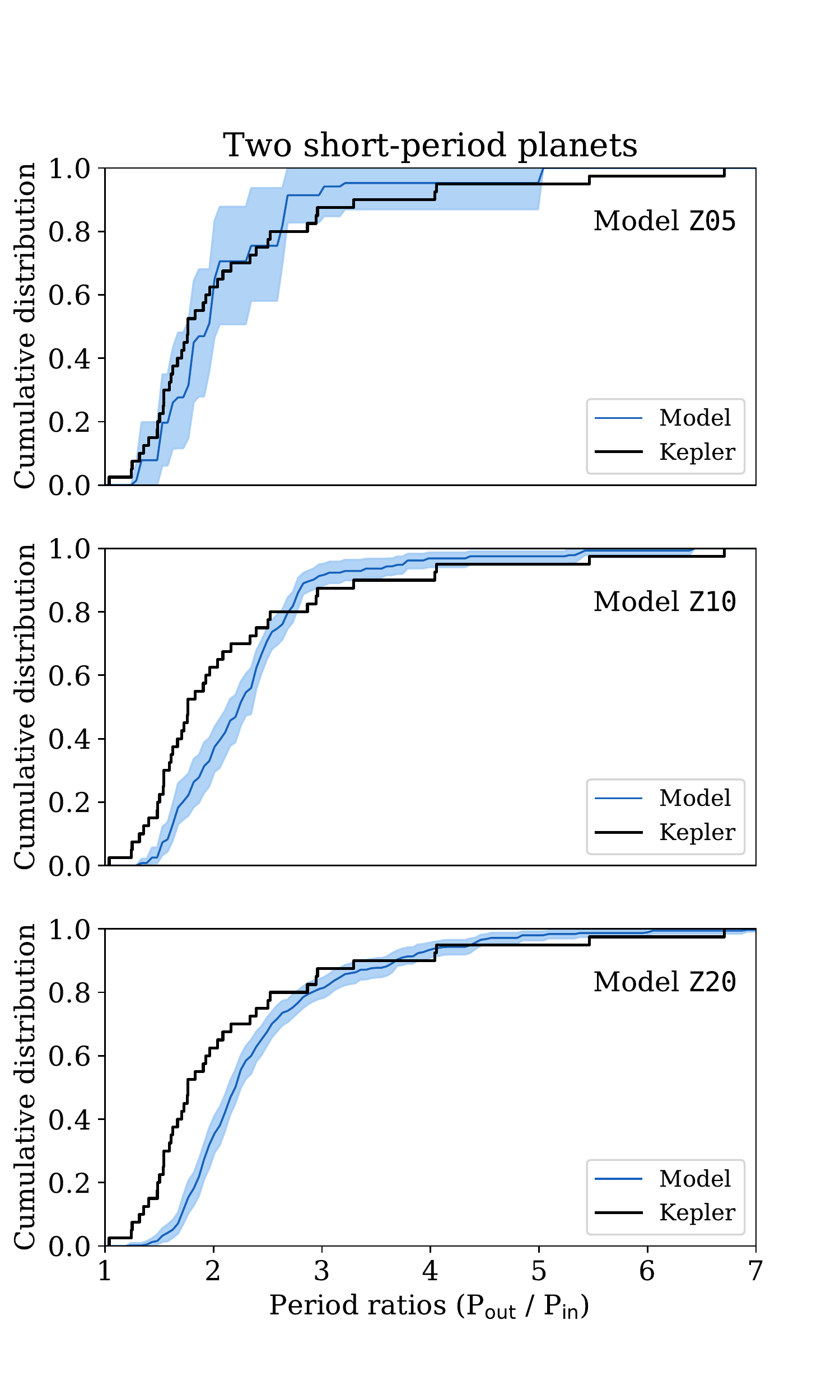}
  \caption{Cumulative distribution of period ratios between short period ($P < 10$ day) planets. \textit{Black}: Kepler catalog of 1-2$R_\oplus$ planets. \textit{Blue}: Simulated planets after applying Kepler detection biases. All short period planets are treated as rocky cores with no gaseous envelope. The light blue bands cover the middle 90\% of bootstrap resamplings, and the solid line is the median. The models have disk metallicities of $Z = 0.5$\%, 1\%, and 2\% respectively.}
  \label{fig:CDF_SS}
\end{figure}

\begin{figure}
  \includegraphics[width=0.47\textwidth]{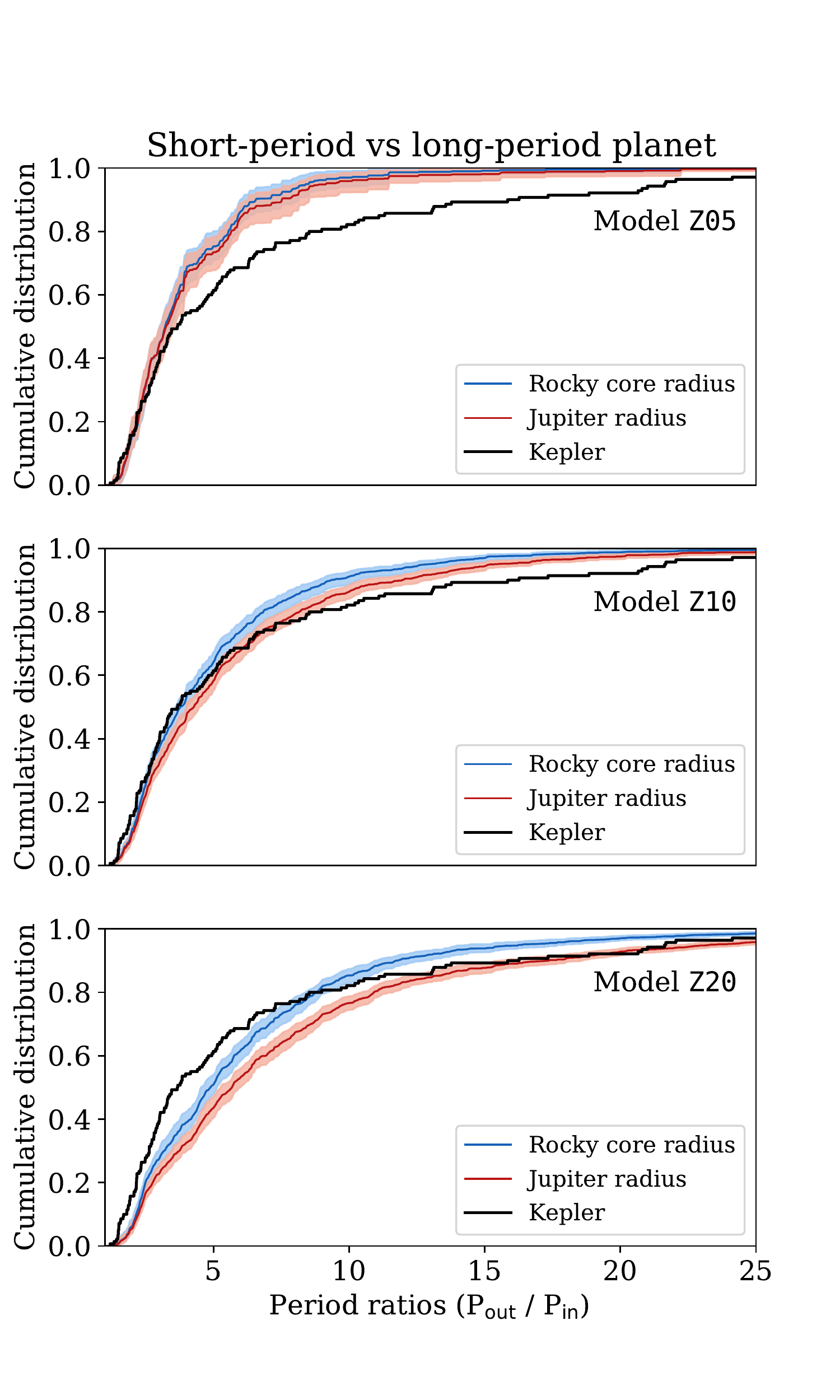}
  \caption{Cumulative distribution of period ratios between one long-period ($P > 10$ day) and one short-periood ($P < 10$ day) planet. \textit{Black}: Kepler catalog of 1-2$R_\oplus$ planets. \textit{Color}: Simulated planets after applying Kepler detection biases. Short period planets are treated as rocky cores. As limiting cases, planets beyond 10 days are treated as either bare rocky cores (blue) or as Jupiter-radius planets (red). The colored bands cover the middle 90\% of bootstrap resamplings. The models have disk metallicities of $Z = 0.5$\%, 1\%, and 2\% respectively.}
  \label{fig:CDF_SL}
\end{figure}

When it comes to the orbital period ratios of short period planets, our simulation results are somewhat ambiguous. Figure \ref{fig:CDF_SS} shows the distribution of period ratios between any two short period planets in the same system. In most of our models, the ``observed'' period ratio between two short period planets is typically larger than observed in the Kepler field. The main exception is Model {\tt Z05}, which also has larger uncertainties due to small number statistics. It is possible that the apparently large period ratios might be due to higher mutual inclinations between planets, as that might make neighbouring planets less likely to both transit.

Looking at more distant neighbours, Figure \ref{fig:CDF_SL} shows the period ratios between any short period planet and any planet with $10\;{\rm d} < P < 100\;{\rm d}$. While none of the models fit the Kepler data perfectly, it seems likely that some combination of these models (i.e. a distribution of disk metallicities) might resolve most of the discrepancy.

%
%
\section{Discussion}
\label{sec:discussion}

\subsection{Planet separations}

Perhaps the most notable discrepancy between our simulation results and Kepler observations is in the separations between short period planets (Figure \ref{fig:CDF_SS}) --- our simulations predict fewer planet pairs with period ratios $P_{\rm out}/P_{\rm in} \lesssim 2$ than is observed in the Kepler catalogue. We have identified three possible effects that might be responsible for this discrepancy:

\begin{figure}
  \includegraphics[width=0.47\textwidth]{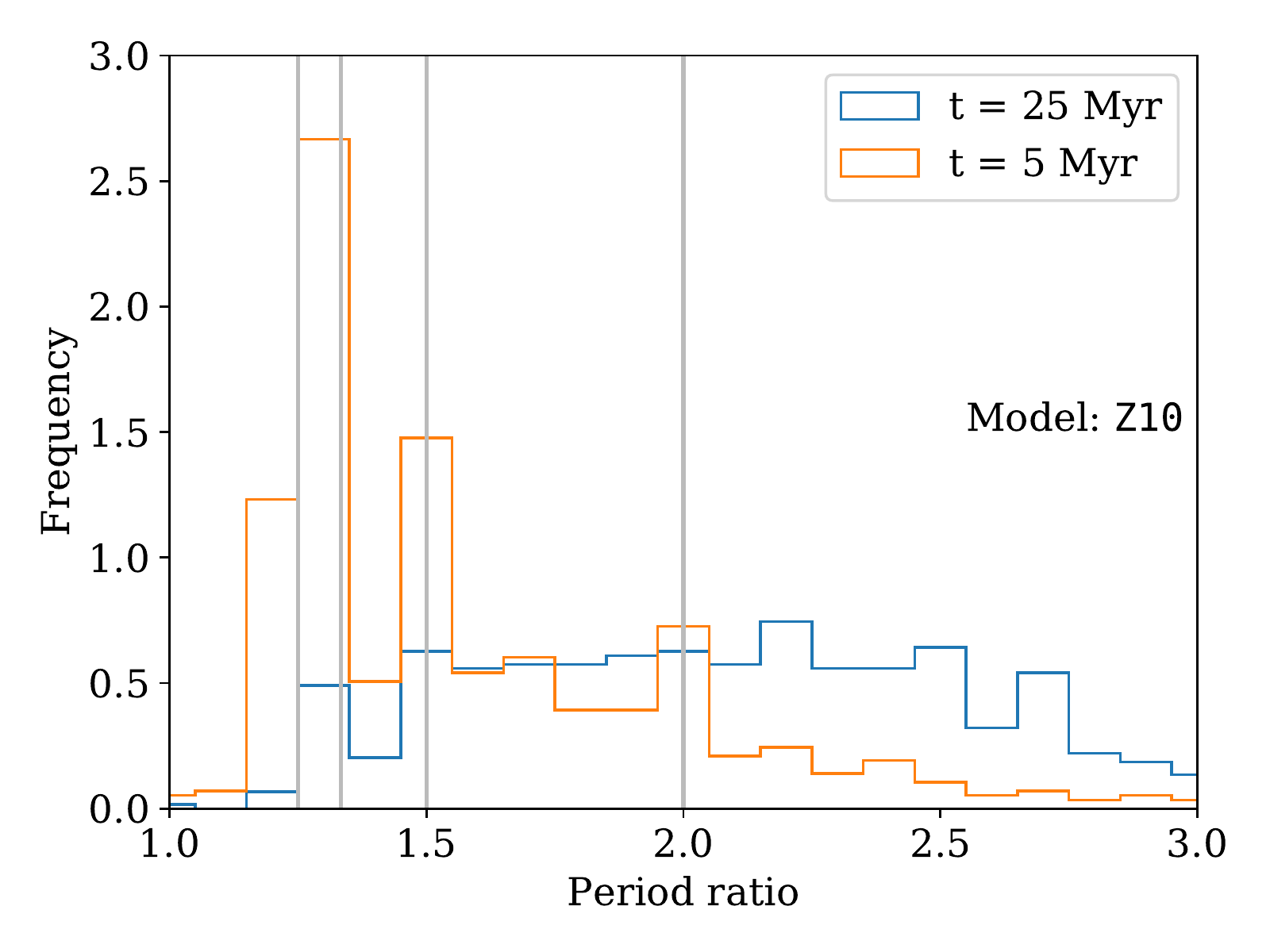}
  \caption{Histogram of the period ratio distribution of adjacent planets at two points in time. The disk lifetime is 5.1 Myr. The grey lines show the locations of the 2:1, 3:2, and 4:3 mean motion resonances. At $t = 5$ Myr (orange) many planet pairs are in or near mean motion resonances. At $t = 25$ Myr (blue) most systems have had collisions, leading to larger separations. This plot does not take observational effects into account. Those effects cause a small tilt toward smaller period ratios.}
  \label{fig:chain}
\end{figure}

\begin{figure}
  \includegraphics[width=0.47\textwidth]{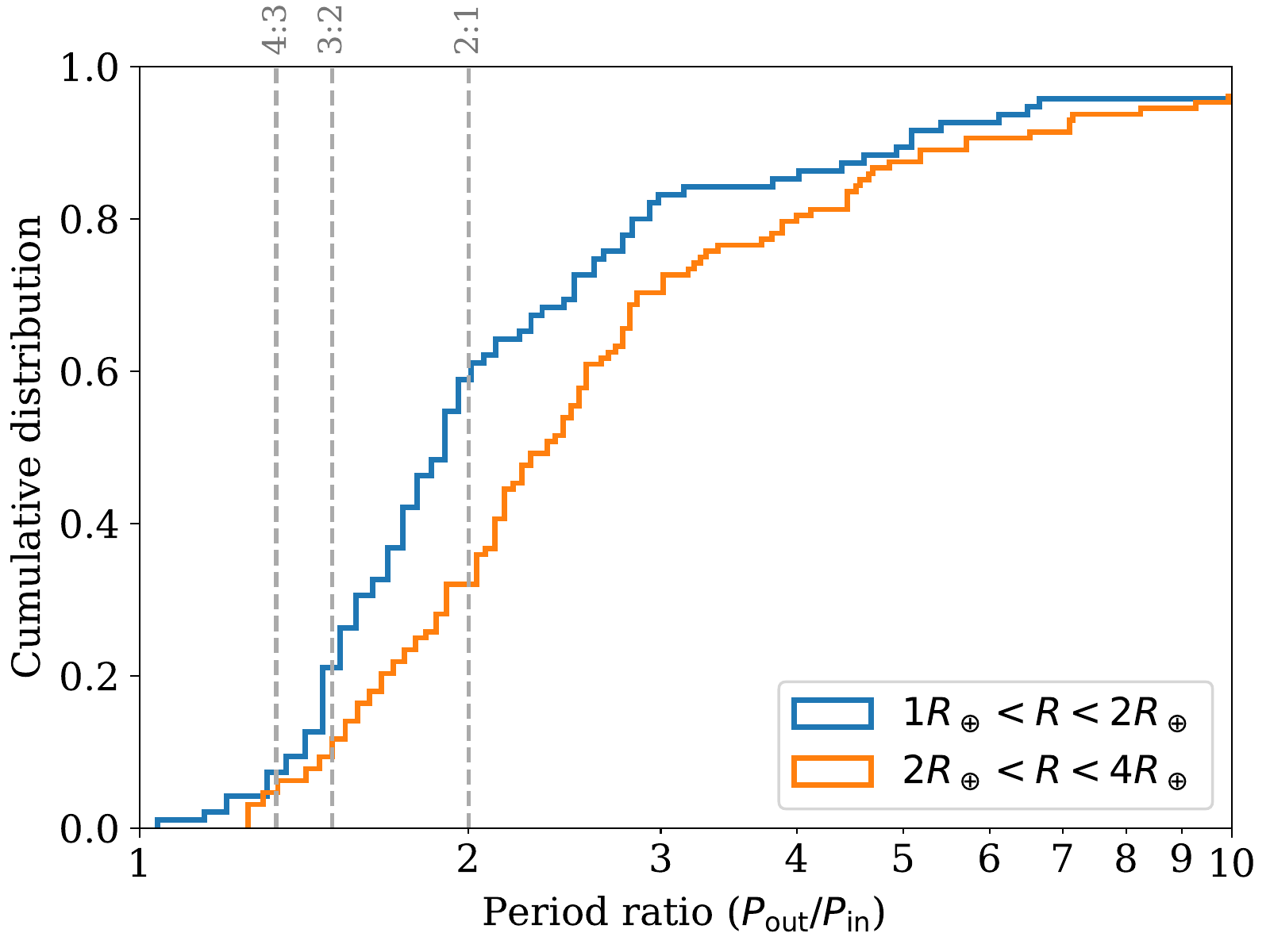}
  \caption{Distribution of period ratios for small Kepler planets with $P < 200$ days, inside two radius bins. Period ratios less than two preferentially come from planetary systems with small radius planets. This may be due to a combination of small planets being more stable at small orbital periods, or more massive planets preferentially being trapped at the 2:1 mean motion resonance.}
  \label{fig:kepler-period-ratios}
\end{figure}

\begin{itemize}
\item One possible explanation is that too many of our systems experience a dynamical instability after the disk dissipates. In dynamical instability, the gravitational interactions between planets cause the orbits to evolve chaotically, leading to orbit crossing and giant impacts. The last two frames of Figure \ref{fig:snapshots} illustrate this concept. After these collisions, the system is left with a smaller number of more widely separated planets. Figure \ref{fig:chain} shows the distribution of period ratios just before the disk dissipates ($t = 5$ Myr) and long after the disk dissipates ($t = 25$ Myr). At 5 Myr, most planet pairs are at or near mean motion resonances and have small period ratios. Over the next 20 Myr, most simulations have a dynamical instability that breaks these resonant chains and causes planets to collide. If our runs overestimate the frequency of instabilities, they may underestimate the number of systems with small period ratios.

\item A related possibility is that there is a correlation between the frequency of dynamical instabilities and some other property of the planetary system, like the masses of the planets. Our simulations under-produce Earth-mass planets (Figure \ref{fig:R_vs_P}) and it stands to reason that low-mass systems should be stable at closer orbital separations. Indeed, as shown in Figure \ref{fig:kepler-period-ratios}, Kepler systems with smaller planets seem to prefer smaller period ratios.

\item Finally, it is possible that there is a correlation with planet mass that is unrelated to dynamical instabilities. Perhaps the most natural alternative is that more massive planets may be more likely to be trapped at the 2:1 mean motion resonance.
\end{itemize}

Future work should investigate how the dynamical history of planetary systems determines the observed distribution of period ratios. In addition, it will be important to explore possible connections between disk properties, such as the total amount of mass in embryos, and the dynamical history of the resulting planetary system.

\subsection{Ultra short period planets}

Ultra short period planets are generally defined as planets with periods less than 1 or 2 days. As Figure \ref{fig:R_vs_P} shows, our simulations generally struggle to produce planets with periods shorter than two days, and none of our simulations produced planets with periods less than a day. There are a couple of factors that we did not take into account that may be responsible for USPs:

\begin{itemize}
\item All of our simulations had the inner edge of the disk at 10 days. But Figure \ref{fig:intro} shows that there is in fact a distribution of stellar rotation rates (and hence, co-rotation radii) all the way down to about 1 day. It is possible that the population of USPs with $P < 1$ day come from the minority of disks where the inner edge is at $P \sim 1$ day.

\item We also did not take into account the effect of stellar tides. Stellar tides become efficient at very short orbital periods ($P \sim 1$ day) and some authors have suggested that they may be responsible for the population of USPs \citep[e.g.][]{Lee_2017}.
\end{itemize}

Future investigations should investigate how these factors couple with the migration of resonant chains. Stellar tides pose a difficult challenge because the strength of the tides is not well understood. The tide raised by a planet on its host star causes the planets semimajor axis to decay at a rate

\begin{equation}
    \dot{a} =   \frac{-9}{2}
                a^{-11/2}G^{1/2}
                \frac{mR_\star^5}{M_\star^{1/2}Q_\star'},
\end{equation}
where $m$ is the planet mass, $M_\star$ and $R_\star$ are the stellar mass and radius, and $Q_\star'$ is the effective tidal quality factor \citep[e.g.][]{Goldreich_1966}. Estimates of the quality factor $Q_\star'$ vary by several orders of magnitude and depend strongly on the assumed properties of the planet and forcing period \citep[e.g.][]{Matsumura_2008,Schlaufman_2010,Penev_2012}.

%
%
\section{Conclusion}
\label{sec:conclusion}

The magnetic coupling between protoplanetary disks and their host stars implies that most protoplanetary disks are likely to be truncated at the point of co-rotation with the star's rotational period, or at about 10 days. We propose that short period planets ($P < 10$ days) may form through the Type-I migration of a resonant chain of planets. The process of super-Earth formation from a population of planetesimals and embryos naturally leads to the formation of long chains of planets that are locked into mutual mean motion resonances \citep[e.g.][]{Terquem_2007,Cossou_2014,Izidoro_2017,Carrera_2018}. When a migrating chain reaches the inner edge of the disk, the innermost planet experiences zero or even outward torques, but the aggregate inward torque from the outer planets in the chain force all planets into smaller orbital periods.

After a careful treatment of selection biases, using a forward model of the transit probabilities, and the detection efficiency of the Kepler pipeline (Sections \ref{sec:intro:obs} and \ref{sec:methods:obs}), we showed that the ``migrating chain'' model predicts period distributions that are consistent with those observed in the Kepler field (Figure \ref{fig:freq_vs_P}). Therefore, we conclude that a majority of short period planets may well be produced through the migration of a resonant chain of planets. The observed period ratios between short period planets and planets with $P > 10$ days were also broadly consistent with observation, but the exact distribution depended greatly on the disk properties.

The main area where the model needs improvement is that the model predicts slightly larger period ratios between short period planets than what is observed, and the models generally struggle to produce ultra short period planets. We speculate that the observed period ratios may be partially tied to the frequency of dynamical instabilities. As for the under-production of ultra short period planets, we suspect that stellar tides, and the fact that some disks probably have an inner edge closer to $P \sim 1$ day, probably play a role.

Our results provide valuable insight into the planet formation process, and suggests that resonance locks, migration, and dynamical instabilities play important roles the the formation and evolution of short period planets.

%
%
\section*{Acknowledgements}
D.C.\,acknowledges Angie Wolfgang who provided helpful discussions and insight on modelling transit detection biases. D.C.'s research was supported by an appointment to the NASA Postdoctoral Program within NASA's Nexus for Exoplanet System Science (NExSS), administered by Universities Space Research Association under contract with NASA. E.B.F.\,acknowledges support from NASA Exoplanet Research Program award NNX15AE21G. The results reported herein benefitted from collaborations and/or information exchange within NASA's Nexus for Exoplanet System Science (NExSS) research coordination network sponsored by NASA's Science Mission Directorate. The Center for Exoplanets and Habitable Worlds is supported by the Pennsylvania State University, the Eberly College of Science, and the Pennsylvania Space Grant Consortium. We gratefully acknowledge support from NSF grant MRI-1626251. This research or portions of this research were conducted with Advanced CyberInfrastructure computational resources provided by The Institute for CyberScience at The Pennsylvania State University (\texttt{http://ics.psu.edu}), including the CyberLAMP cluster supported by NSF grant MRI-1626251.

%
%
\bibliographystyle{mnras}
\bibliography{references}

%
%
\appendix

%
%
\section{Disk torques}
\label{app:torques}

This appendix is a slightly condensed rewrite of Appendix A of \citet{Carrera_2018} with the correction of two minor typos. Following \citet{Paardekooper_2010,Paardekooper_2011}, all the formulas below include a smoothing of the planet potential with a smoothing length of $b = 0.4h$ where $h = H/r \approx 0.05$. In Type-I migration, a planet experiences a negative Lindblad torque $\Gamma_{\rm L}$ and a positive co-rotation torque $\Gamma_{\rm C}$. The total torque on the planet is given by

\begin{equation}\label{eqn:Gamma_tot}
	\Gamma_{\rm tot} = \Gamma_{\rm L} \Delta_{\rm L}
    				 + \Gamma_{\rm C} \Delta_{\rm C}
\end{equation}

The expressions for $\Delta_{\rm L}$ and $\Delta_{\rm C}$ were produced by \citet{Cresswell_2008,Coleman_2014,Fendyke_2014} and collected by \citet{Izidoro_2017},

\begin{eqnarray}\nonumber
    \Delta_{\rm L}
 	&=&
    \left[
    P_{\rm e} + \frac{P_{\rm e}}{|P_{\rm e}|} \times
    \left\lbrace
    	0.07 \left( \frac{i}{h}\right)
    	+ 0.085\left( \frac{i}{h}\right)^4\right.\right.\\
    && \left.\left.
        - 0.08\left(  \frac{e}{h} \right) \left( \frac{i}{h} \right)^2
	\right\rbrace
    \right]^{-1}\\
	\Delta_{\rm C}
    &=&	\exp\left( \frac{e}{e_{\rm f}} \right)
    	\left\lbrace
        	1 - \tanh\left(\frac{i}{h} \right)
        \right\rbrace
\end{eqnarray}

where $e$ and $i$ are the planet orbital eccentricity and inclination, where $e_{\rm f} = 0.5h + 0.01$, and 

\begin{equation}
	P_{\rm e}
    = \frac{
    	1 + \left( \frac{e}{2.25h}\right)^{1.2} + \left( \frac{e}{2.84h}\right)^6
	}{
    	1 - \left( \frac{e}{2.02h}\right)^4
	}.
\end{equation}

The expressions for $\Gamma_{\rm L}$ and $\Gamma_{\rm C}$ were derived by \citet{Paardekooper_2010,Paardekooper_2011}

\begin{eqnarray}
	\frac{\Gamma_{\rm L}}{\Gamma_0/\gamma_{\rm eff}}
    &=& (-2.5 -1.7\beta + 0.1x),\\
    \nonumber
	\frac{\Gamma_{\rm C}}{\Gamma_0/\gamma_{\rm eff}}
    &=&   1.1\left( \frac{3}{2}-x\right) F(p_\nu)~G(p_\nu)\\
    \nonumber
	&+&   0.7\left( \frac{3}{2}-x\right) (1 - K(p_\nu)) \\
    &+&     \frac{7.9 \xi}{\gamma_{\rm eff}} F(p_\nu) F(p_\chi)
            \sqrt{G(p_{\rm \nu}) G(p_\chi)}\\
	&+&     \left( 2.2 - \frac{1.4}{\gamma_{\rm eff}}\right)\xi
    		\sqrt{(1 - K(p_\nu))~(1 - K(p_\chi)}\nonumber
\end{eqnarray}

where $\xi = \beta - (\gamma -1)x$ is the negative of the entropy slope, $\gamma = 1.4$ is the adiabatic index, $x$ is the negative of the surface density profile, $\beta$ is the temperature gradient, and $\Gamma_0$ is a scaling factor,

\begin{equation}
	x = - \frac{\partial {\rm ln} ~\Sigma_{\rm gas}}{\partial {\rm ln}~r},
	~~~
    \beta = - \frac{\partial {\rm ln} ~T}{\partial {\rm ln}~r}.
	~~~
	\Gamma_0 = \left(\frac{q}{h}\right)^2 \Sigma_{\rm gas} r^4 \Omega_k^2,
\end{equation}
and $q$ is the planet-star mass ratio. Next, we define the effective adiabatic index $\gamma_{\rm eff}$

\begin{eqnarray}
	\gamma_{\rm eff}
    &=&
    \frac{2~Q \gamma}{
    	\gamma Q + \frac{1}{2}
        \sqrt{2 \sqrt{( \gamma^2Q^2+1)^2-16 Q^2(\gamma-1)}+2 \gamma^2 Q^2 - 2}
    },\\
	Q
    &=&
    \frac{2 \chi}{3 h^3 r^2 \Omega_k},\\
	\chi
    &=&
    \frac{16~\gamma~(\gamma -1)~\sigma T^4}{3 \kappa~\rho^2~(hr)^2~\Omega_k^2},
\end{eqnarray}
where $\rho$ is the gas volume density, $\kappa$ is the opacity and $\sigma$ is the Stefan-Boltzmann constant. Next, $p_\nu$ and $p_\chi$ are parameters that govern the viscous saturation and the thermal saturation respectively. They are defined in terms of the non-dimensional half-width of the horseshoe region ${x_{\rm s}}$,

\begin{equation}
	p_\nu = \frac{2}{3}\sqrt{\frac{r^2\Omega_k}{2\pi\nu}x_s^3},
	~~~
	p_\chi = \sqrt{\frac{r^2\Omega_k}{2\pi\chi}x_s^3},
	~~~
	x_{\rm s} = \frac{1.1}{\gamma_{\rm eff}^{1/4}}\sqrt{\frac{q}{h}}.
\end{equation}

Finally, we give the expression for functions $F$, $G$, and $K$,

\begin{eqnarray}
  F(p)
  &=& \frac{1}{1+ \left( \frac{p}{1.3}\right)^2}\\
  G(p)
  &=&
    \left\{
        \begin{array}{ll}
            \frac{16}{25}
            \left( \frac{45\pi}{8}\right)^{3/4} p^{3/2}
            & {\rm if}~~ p < \sqrt{\frac{8}{45\pi}}\\
            1 - \frac{9}{25}
            \left( \frac{8}{45\pi}\right)^{4/3} p^{-8/3}
            & {\rm otherwise}
        \end{array}
    \right.\\
  K(p)
  &=&
    \left\{
        \begin{array}{ll}
            \frac{16}{25}
            \left( \frac{45\pi}{28} \right)^{3/4} p^{3/2}
            & {\rm if}~~ p < \sqrt{\frac{28}{45\pi}}\\
            1 - \frac{9}{25}
            \left( \frac{28}{45\pi} \right)^{4/3} p^{-8/3}
            & {\rm otherwise}\label{eqn:Kp}
        \end{array}
    \right.
\end{eqnarray}

Equations \ref{eqn:Gamma_tot}-\ref{eqn:Kp} are enough to compute the total torque $\Gamma_{\rm tot}$. To implement disk torques as an external force in an N-body code, we follow \citet{Papaloizou_2000} and \citet{Cresswell_2008} in defining the planet migration timescale as $t_{\rm mig} =- L / \Gamma_{\rm tot}$, where $L$ is the planet's orbital angular momentum. The force term becomes,

\begin{equation}\label{eqn:a_mig}
	\mathbf{a}_{\rm mig} = -\frac{\mathbf{v}}{t_{\rm mig}},
\end{equation}
where $\mathbf{v}$ is the planet's instantaneous velocity. To implement eccentricity and inclination damping, we need their damping timescales ${\rm t_e}$ and ${\rm t_i}$. Expressions for these timescales have been derived by \citet{Papaloizou_2000} and \citet{Tanaka_2004}, and were later modified by \citet{Cresswell_2006,Cresswell_2008},

\begin{eqnarray}
	t_e
    &=&
    \frac{t_{\rm wave}}{0.780}
    \left[
    	1-0.14\left(\frac{e}{h}\right)^2 
        + 0.06\left(\frac{e}{h}\right)^3\right.\\
    &+&\left.
        0.18\left(\frac{e}{h}\right)\left(\frac{i}{h}\right)^2
    \right]\nonumber\\
	t_i
    &=&
    \frac{t_{\rm wave}}{0.544}
    \left[
    	1-0.3\left(\frac{i}{h}\right)^2 
        + 0.24\left(\frac{i}{h}\right)^3\right.\\
    &+&\left.
        0.14\left(\frac{e}{h}\right)^2\left(\frac{i}{h}\right)
    \right]\nonumber\\
	t_{\rm wave}
    &=&
    \left(\frac{M_{\odot}}{m}\right)
    \left(\frac{M_{\odot}}{\Sigma_{\rm gas} a^2}\right)
    h^4 \Omega_k^{-1},
\end{eqnarray}
where $m$ and $a$ are the planet's mass and semimajor axis. Similar to Equation \ref{eqn:a_mig}, we implement eccentricity damping and inclination damping, respectively, as the forces

\begin{equation}
	\mathbf{a}_e = -2~\frac{(\mathbf{v \cdot r})\mathbf{r}}{r^2 t_e},
	~~~
	\mathbf{a}_i = -\frac{v_z}{t_i}\mathbf{k}
\end{equation}

where $\mathbf{r}$ and $\mathbf{v}$ are the position and velocity vectors of the planet, $v_z$ is the $z$ component of the planet's velocity, and $\mathbf{k}$ is the unit vector in the $z$ direction.

%
%
\bsp	
\label{lastpage}
\end{document}